\documentclass[sigconf]{aamas} 

\usepackage{balance} 
\usepackage{graphicx}
\usepackage{stfloats}

\newtheorem{definition}{Definition}

\setcopyright{none}
\acmConference[ALA '24]{Proc.\@ of the Adaptive and Learning Agents Workshop (ALA 2024)}
{May 6-7, 2024}{Online, \url{https://ala2024.github.io/}}{Avalos, Müller, Wang, Yates (eds.)}
\copyrightyear{2024}
\acmYear{2024}
\acmDOI{}
\acmPrice{}
\acmISBN{}
\title{Multi-Agent Synchronization Tasks}

\author{Rolando Fernandez}
\affiliation{
  \institution{DEVCOM Army Research Laboratory\\ The University of Texas at Austin}
  \city{Austin, TX}
  \country{United States}}
\email{rfernandez@utexas.edu}

\author{Garrett Warnell}
\affiliation{
  \institution{DEVCOM Army Research Laboratory\\ The University of Texas at Austin}
  \city{Austin, TX}
  \country{United States}}
\email{garrett.a.warnell.civ@army.mil}

\author{Derrik E. Asher}
\affiliation{
  \institution{DEVCOM Army Research Laboratory}
  \city{Adelphi, MD}
  \country{United States}}
\email{derrik.e.asher.civ@army.mil}

\author{Peter Stone}
\affiliation{
  \institution{The University of Texas at Austin \\ Sony AI}
  \city{Austin, TX}
  \country{United States}}
\email{pstone@cs.utexas.edu}

\begin{abstract}
In multi-agent reinforcement learning (MARL), coordination plays a crucial role in enhancing agents' performance beyond what they could achieve through cooperation alone. The interdependence of agents' actions, coupled with the need for communication, leads to a domain where effective coordination is crucial. In this paper, we introduce and define \textit{Multi-Agent Synchronization Tasks} (MSTs), a novel subset of multi-agent tasks. We describe one MST, that we call \textit{Synchronized Predator-Prey}, offering a detailed description that will serve as the basis for evaluating a selection of recent state-of-the-art (SOTA) MARL algorithms explicitly designed to address coordination challenges through the use of communication strategies. Furthermore, we present empirical evidence that reveals the limitations of the algorithms assessed to solve MSTs, demonstrating their inability to scale effectively beyond 2-agent coordination tasks in scenarios where communication is a requisite component. Finally, the results raise questions about the applicability of recent SOTA approaches for complex coordination tasks (i.e. MSTs) and prompt further exploration into the underlying causes of their limitations in this context.
\end{abstract}


\newcommand{\BibTeX}{\rm B\kern-.05em{\sc i\kern-.025em b}\kern-.08em\TeX}

\begin{document}


\pagestyle{fancy}
\fancyhead{}


\maketitle 


\section{Introduction}

Coordination is essential in fully cooperative environments where agents on a team must work together to achieve a common goal. In this work, we define \textit{Coordination} as the ability of agents to align their actions with their partners and \textit{Cooperation} as the ability of agents to work together to achieve a shared objective (e.g., by sharing information about their actions to gain mutual benefit) \cite{fernandez2021emergent, asher2019multi}. Together \textit{Coordination} and \textit{Cooperation} allow agents to communicate and collaborate to make joint decisions and take collective actions that lead to the best outcomes for the team. 

In multi-agent reinforcement learning (MARL), coordination plays a crucial role in enhancing agents' performance beyond what they could achieve through cooperation alone. This performance improvement is achieved by enabling agents to share information and align their actions. Coordination is essential in MARL because it enables teams to work together more effectively than without coordination, especially when the agents' objectives are interdependent or when their actions can impact each other. For example, in a soccer game, players must coordinate their actions to score and defend their goals.

Typically, MARL approaches leverage cooperative domains tailored to facilitate the study of agent interactions and coordination strategies. These cooperative environments share common attributes essential for testing and refining multi-agent systems. Cooperative domains are characterized by agents working toward shared goals, where the success of each agent is intertwined with the overall team objective. The interdependence of agents' actions, coupled with the need for communication, leads to a domain where effective coordination is crucial. Tasks within these domains are deliberately complex, often featuring dynamic elements and uncertainties that demand adaptive coordination strategies. Some examples of these types of domains are predator-prey, stag hunt, and StarCraft, among others \cite{lowe2017multi, rashid2020monotonic, bohmer2020dcg, li2021dicg, kortvelesy2022qgnn}. Additionally, one-shot game domains have been employed to analyze various MARL methods, contributing to a comprehensive understanding of the capabilities and limitations of the methods \cite{castellini2019representational, kortvelesy2022qgnn}. These environments provide diverse challenges for evaluating the robustness and adaptability of MARL approaches.

In this paper, the focus is directed towards Multi-agent Synchronization Tasks (MSTs), which are a novel subset of cooperative multi-agent domains. MSTs necessitate communication strategies and precise timing for agents to align their actions and accomplish a given task. The emphasis on communication underscores the importance of deliberate information exchange among agents, highlighting the need for a well-defined communication framework to achieve coordination. The requirement for precise timing further accentuates the intricacies of these domains, where the synchronization of agents' actions becomes a critical factor in the successful execution of tasks. By honing in on MSTs, the paper aims to contribute insights into the challenges and strategies associated with explicit communication and precise timing in multi-agent systems, thereby advancing our understanding of cooperative behaviors in complex environments. 

We will describe one such domain, that we call \textit{Synchronized Predator-Prey}, offering a detailed description that will serve as the basis for evaluating a selection of recent MARL algorithms explicitly designed to address coordination challenges through the use of communication strategies. Our analysis will shed light on how the existence of penalties for mis-timed (unsychrnonized) actions, manifested as negative rewards, influences the need for communication between agents (i.e. to avoid relative overgeneralization \cite{panait2006biasing}). Furthermore, we will present empirical evidence that reveals the limitations of the assessed MARL algorithms, demonstrating their inability to scale effectively beyond two-agent coordination tasks in scenarios where communication is a requisite component. This investigation aims to provide an understanding of the interplay between communication, penalties, and scalability within cooperative multi-agent systems, offering valuable insights into the performance and limitations of contemporary MARL approaches in such contexts.

\section{Related Work}\label{related_work}

We now discuss the relationship between Multi-agent Synchronization Tasks (MSTs) and the MARL literature. We will discuss multi-agent synchronization tasks in the context of several well-known MARL domains. Additionally, we will discuss several current state-of-the-art (SOTA) MARL algorithms that we will evaluate on synchronization tasks.

\subsection{Domains}

In this work, we are concerned with a particular type of multi-agent problem that requires agents to align their actions to succeed in a given task. Such tasks have been explored in the MARL literature via established domains such as Stag Hunt and Predator-Prey Pursuit games \cite{skyrms2001stag, morice2013predator}. More recent approaches have also employed the StarCraft Multi-Agent Challenge (SMAC) domain and one-shot (i.e., non-sequential) domains \cite{castellini2019representational, kortvelesy2022qgnn}. However, none of these domains highlights the need for precise synchronization among agents (see Section 3), and so we will introduce here a new domain that we refer to as \emph{Synchronized Predator-Prey}.

\subsection{Algorithms}

We now briefly describe existing algorithmic approaches designed to engender coordination among agents in MARL. For more detailed information, refer to the methods/algorithms sections of the cited works. We focus on the use of communication methods to solve multi-agent synchronization tasks; specifically on the communication method called \textit{Coordination Graphs} (CG), a graph representation of the communication structure between agents in a multi-agent system. While CGs were first combined with RL in the work of Guestrin et al. \cite{guestrin2002cg}, there have been several recent advances in this space.

\noindent \textbf{DCG.} Building on the work of Guestrin, Bohmer et al. \cite{bohmer2020dcg} proposed Deep Coordination Graphs (DCG), a MARL algorithm based on Deep Q Networks (DQN) that utilized CGs to coordinate greedy action selection between connected agents in a graph. Instead of the Variable Elimination scheme, they use the Max-Plus anytime messaging scheme \cite{kok2006MaxPlus}, which trades the exact optimal joint action for an approximation that allows real-time action selection. Though this approach overcomes the real-time limitations of previous works, it still maintains the use of pre-defined coordination graphs with fixed topologies.

\noindent \textbf{DICG.} To overcome the limitations of using pre-defined fixed coordination graphs, Li et al. \cite{li2021dicg} proposed Deep Implicit Coordination Graphs (DICG). This method is a policy-based algorithm that allows implicit coordination graphs to be constructed implicitly at each timestep given the state of the environment. This new algorithm introduced two main changes to the previous work: (1) a Graph Neural Network (GNN) \cite{kipf2016GNN} is used to handle message passing between agents, and (2) an attention network is introduced, which implicitly creates the coordination graph at each timestep. 

\noindent \textbf{QGNN.} Similar to the DICG application of GNNs, Kortvelesy et al. \cite{kortvelesy2022qgnn} proposed a DQN-based value function factorization method called QGNN that uses a model and mixer framework akin to QMIX \cite{rashid2020monotonic}, but with a GNN \cite{kipf2016GNN} embedded in the model network. To summarize, QGNN introduces three main changes to the previous work: (1) it utilizes a model and mixer framework like QMIX, (2) a GNN \cite{kipf2016GNN} is embedded in the model network, and (3) the QGNN mixer is introduced to compute the monotonic joint-action value using a custom pooling operation. 

\section{Multi-Agent Synchronization Tasks}\label{sec:msts}

In this paper, we are concerned with a particular subset of multi-agent tasks that we call multi-agent synchronization tasks (MSTs). In this section, we provide a precise technical definition of MSTs, and also discuss a sample domain that we have designed that serves as an exemplar.

\subsection{Background}

We define MSTs as a subset of decentralized partially observable Markov decision processes (Dec-POMDPs) \cite{oliehoek2016concise}. Formally, for $n$ agents, a Dec-POMDP is characterized by a tuple:

\begin{equation*}
\mathcal{M}_{DecP} = \langle \mathbb{D}, \mathbb{S}, \mathbb{A}, T, \mathbb{O}, \sigma, R, \gamma \rangle.
\end{equation*}

\noindent
Here, $\mathbb{D} = {1, \ldots, n}$ is the set of $n$ agents and $\mathbb{S}$ denotes a set of states. $\mathbb{A}$ is the set of joint actions $\mathbb{A} = \mathbb{A}^1 \times \ldots \times \mathbb{A}^n$, where $\mathbb{A}^i$ is the set of actions available to agent $i \in \mathbb{D}$, which can be different for each agent. At each timestep $t$, the system is in state $s_t \in \mathbb{S}$, and each agent $i$ takes an action $a^i_t$, leading to one joint action $a_t = (a^1_t, \ldots, a^n_t)$ at each timestep. The resulting next state, $s_{t+1}$, is drawn from the transition function $T(s_{t+1} \mid s_t, a_t)$. The transition yields a shared reward $R_t = R(s_t, a_t)$ with $\gamma \in [0,1)$ denoting the discount factor. Similar to $\mathbb{A}$, $\mathbb{O} = \mathbb{O}^1 \times \ldots \times \mathbb{O}^n$ is the set of joint observations, where $\mathbb{O}^i$ is the set of observations available to agent $i \in \mathbb{D}$. Each agent $i$ observes the state only through a (partial) observation $o^i_t \in \mathbb{O}^i$ drawn from its observation function $\sigma^i(o \mid a_t, s_{t+1})$.

A solution to a Dec-POMDP is typically an optimal policy or a set of optimal policies that specify the actions each agent should execute, considering their individual histories of observations and actions, as informed by the optimal state-action value function ($Q^*$), where 
\begin{align*}
Q^*(b, a) = \sum_{s \in \mathbb{S}} b(s) \left[ R(s, a) + \gamma \sum_{o \in \mathbb{O}} \sigma(o | s, a) \max_{a'} Q^*(b', a') \right].    
\end{align*}
Because the state of the environment is not fully observable to the agents, these policies must be based on beliefs or estimates of the state ($b$), derived from the history of observations and actions. The overarching aim of these policies is to maximize the collective reward over time, optimizing the expected benefit for all agents involved across a potentially infinite or finite horizon. By striving to maximize this collective reward, the solution inherently seeks the most advantageous outcome for the group as a whole, navigating the complexities of partial observability and decentralized decision-making in Dec-POMDP environments.

\subsection{MST Definition}\label{sec:msts_def}

Broadly speaking, MSTs are tasks that exhibit high stakes when it comes to agents synchronizing their actions. We characterize them using per-agent notions of \textit{neutral actions} ($\mathbb{A}^{i, \text{ } neut}$) and \textit{synchronization actions} ($\mathbb{A}^{i, \text{ } sync}$). Neutral actions are standard actions that an agent can take without having to coordinate with others, whereas synchronization actions require precise timing and coordination with other agents to be successful. Intuitively, synchronization actions are special actions that, in at least one circumstance, are optimal, but also carry the risk that, if attempted by some members of the team but not executed by others at the same time, the result is worse than if all members of the team had restricted themselves to neutral actions only.

The formal structure of these tasks involves dividing each agent's possible actions into synchronization and neutral actions, with the total action space for an agent being the union of these two types. This division allows us to categorize joint actions (actions taken by all agents together) into three subsets based on their outcomes in a given state, \textit{neutral} joint actions ($\mathbb{A}^{neut}$), \textit{synchronization-positive} joint actions ($\mathbb{A}^+(s)$), and \textit{synchronization-negative} joint actions ($\mathbb{A}^-(s)$). $\mathbb{A}^{neut}$ consists of joint actions in which every agent opts for a neutral action. These actions are safe, but may not always lead to the best possible outcomes. $\mathbb{A}^+(s)$ is the set of actions that, when executed by the agents, result in a successful synchronization, offering a better reward than any set of neutral actions in the same state. ($\mathbb{A}^-(s)$) is the set of actions where attempts at synchronization fail, leading to outcomes worse than if all agents had chosen neutral actions. For a task to satisfy the definition of an MST, there must exist at least one state such that the sets $\mathbb{A}^+(s)$ and $\mathbb{A}^-(s)$ are not empty.

Moving a large/heavy piece of furniture is an example of an MST where two or more people must coordinate their actions to achieve task success (e.g., avoid hurting themselves and/or damaging the furniture). In this case, the $\mathbb{A}^{i, \text{ } neut}$ actions consist of not lifting the piece of furniture and the $\mathbb{A}^{i, \text{ } sync}$ actions consist of lifting the piece of furniture. The joint actions in $\mathbb{A}^+$ are those such that the two people lift up the piece of furniture simultaneously. Then, the joint actions in $\mathbb{A}^-$ are those such that one person lifts, while the other does not, or such that they both lift, but one person is not next to the furniture, as the lifter could get hurt, and/or the furniture could be damaged. While $\mathbb{A}^{neut}$, are those actions such that no person attempts to lift the piece of furniture, as no one can be injured and the furniture cannot be damaged. From the example, it is clear that actions resulting in successful synchronization are a better choice over neutral actions, while actions that result in unsuccessful synchronization are clearly worse.

More formally, let $(\mathbb{A}^{1, \text{ } sync} , \ldots , \mathbb{A}^{n, \text{ } sync})$ define a partitioning of each agent's action space into $\mathbb{A}^i = \mathbb{A}^{i, \text{ } sync} \text{ } \cup \text{ } \mathbb{A}^{i, \text{ } neut}$ where $\mathbb{A}^{i, \text{ } neut} = \overline{\mathbb{A}^{i, \text{ } sync}}$. Further, let that partitioning define the following three subsets of the joint action space:


\begin{itemize}
    \item[] $Q^* = \mathbb{A}^{1, \text{ } neut} \times \dots \times \mathbb{A}^{n, \text{ } neut}$,
    \newline
    \item[] $\mathbb{A}^+(s) = \Bigg\{
        \text{ } a \text{ } \Bigg\vert \text{ } \begin{array}{l}
        \exists \text{ } i,j \text{ } s.t. \text{ } \begin{array}{l}
                                                        a^i \in \mathbb{A}^{i, \text{ } sync}
                                                        \\
                                                         a^j \in \mathbb{A}^{j, \text{ } sync}
                                                    \end{array}
        \\
        \forall \text{ } a^{neut} \in \mathbb{A}^{neut} \text{ } Q^*(s,a) > Q^*(s,a^{neut})
        \end{array}
        \Bigg\}$,
    \newline and \newline
    \item[] $\mathbb{A}^-(s) = \Bigg\{
        \text{ } a \text{ } \Bigg\vert \text{ } \begin{array}{l}
        \exists \text{ } i \text{ } s.t. \text{ } a^i \in \mathbb{A}^{i, \text{ } sync}
        \\
        \forall \text{ } a^{neut} \in \mathbb{A}^{neut} \text{ } Q^*(s,a^{neut}) > Q^*(s,a)
        \end{array}
        \Bigg\}$.
    \newline 
\end{itemize}

\noindent Note that the definition of $\mathbb{A}^+(s)$ depends on \textit{at least} two agents synchronizing, but does not exclude joint actions where more than two agents synchronize.

\begin{definition}
    \textbf{Multi-agent Synchronization Task}. A multi-agent synchronization task (MST) is a Dec-POMDP where there exists a partitioning $(\mathbb{A}^{1, \text{ } sync}, \ldots, \mathbb{A}^{n, \text{ } sync})$ and at least one $s \in \mathbb{S} $ such that $\mathbb{A}^+(s) \neq \varnothing$ and $\mathbb{A}^-(s) \neq \varnothing$.    
\end{definition}

The presence of elements in the set $\mathbb{A}^+(s)$ indicates that, when the system is in state $s$, there is at least one sub-team of agents for which the execution of synchronization actions ($\mathbb{A}^{i, \text{ } sync}$) is the optimal action. Conversely, non-empty $\mathbb{A}^-(s)$ indicates that, for the same state, there are scenarios where at least one agent attempting a synchronization action results in a catastrophic failure in the sense that, if all agents had stuck to neutral actions, the outcome would have been better. Both sets being non-empty for the same $s$ creates the condition for a high-stakes scenario: it is possible for agents to either achieve the highest level of success through the correct application of synchronization actions, but it is also possible that trying to execute a synchronization action can lead to a worse outcome than trying no synchronization actions at all. This duality underlines the critical importance of precise coordination and timing among agents in MSTs.

\subsection{Example MST: Synchronized Predator-Prey}

We now describe \textit{Synchronized Predator-Prey}, a new variant of the classical Predator-Prey domain that satisfies the definition of an MST.

\begin{figure}[h]
  \centering
  \includegraphics[width=0.9\linewidth]{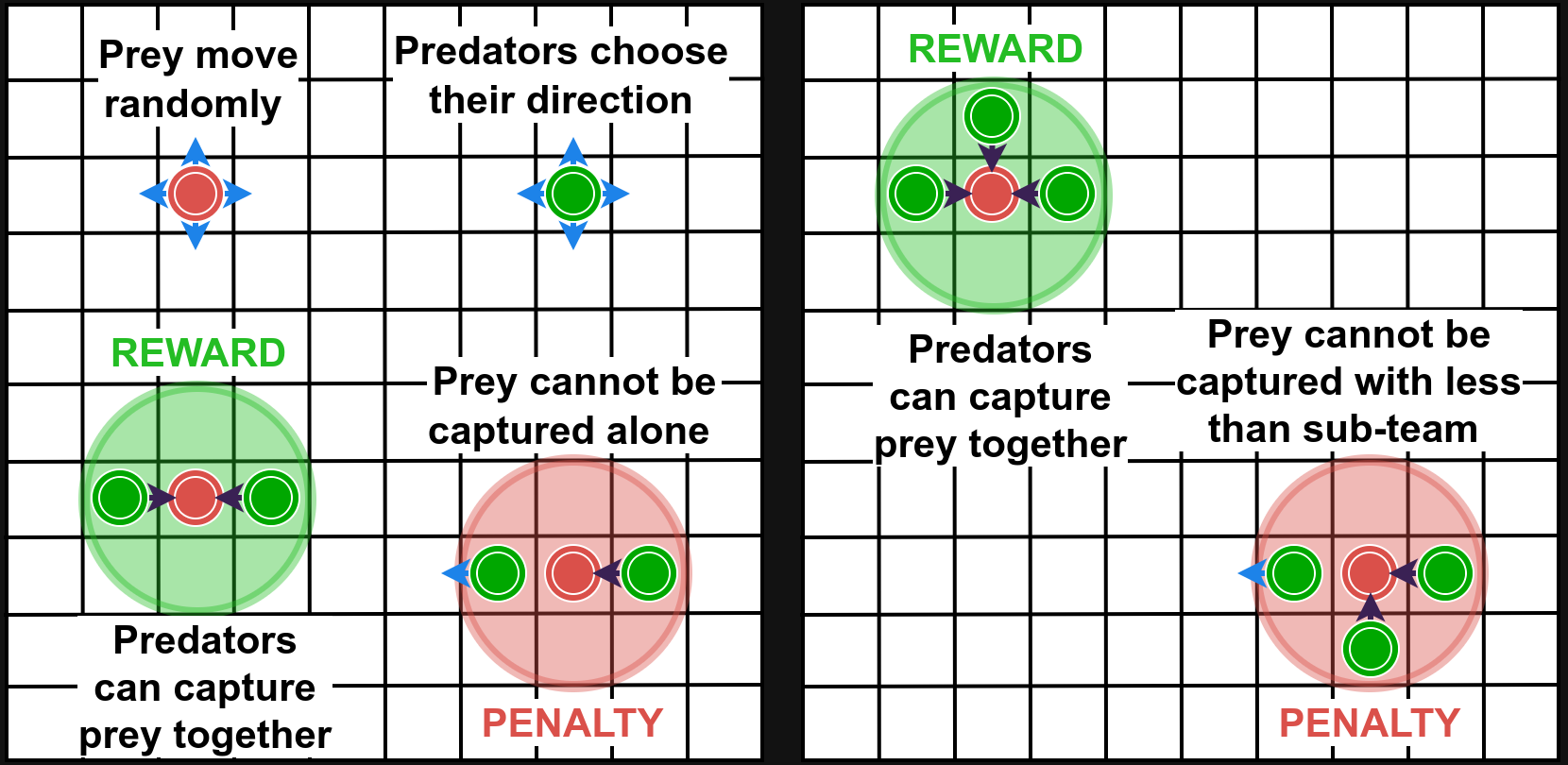}
  \caption{Synchronized Predator-Prey Task. Blue arrows denote movement (neutral) actions and purple arrows denote capture (synchronization) actions.}
  \label{fig:synchronized_pred-prey_w_3_catcher}
\end{figure}

\begin{figure}[h]
  \centering
  \includegraphics[width=1.05\linewidth]{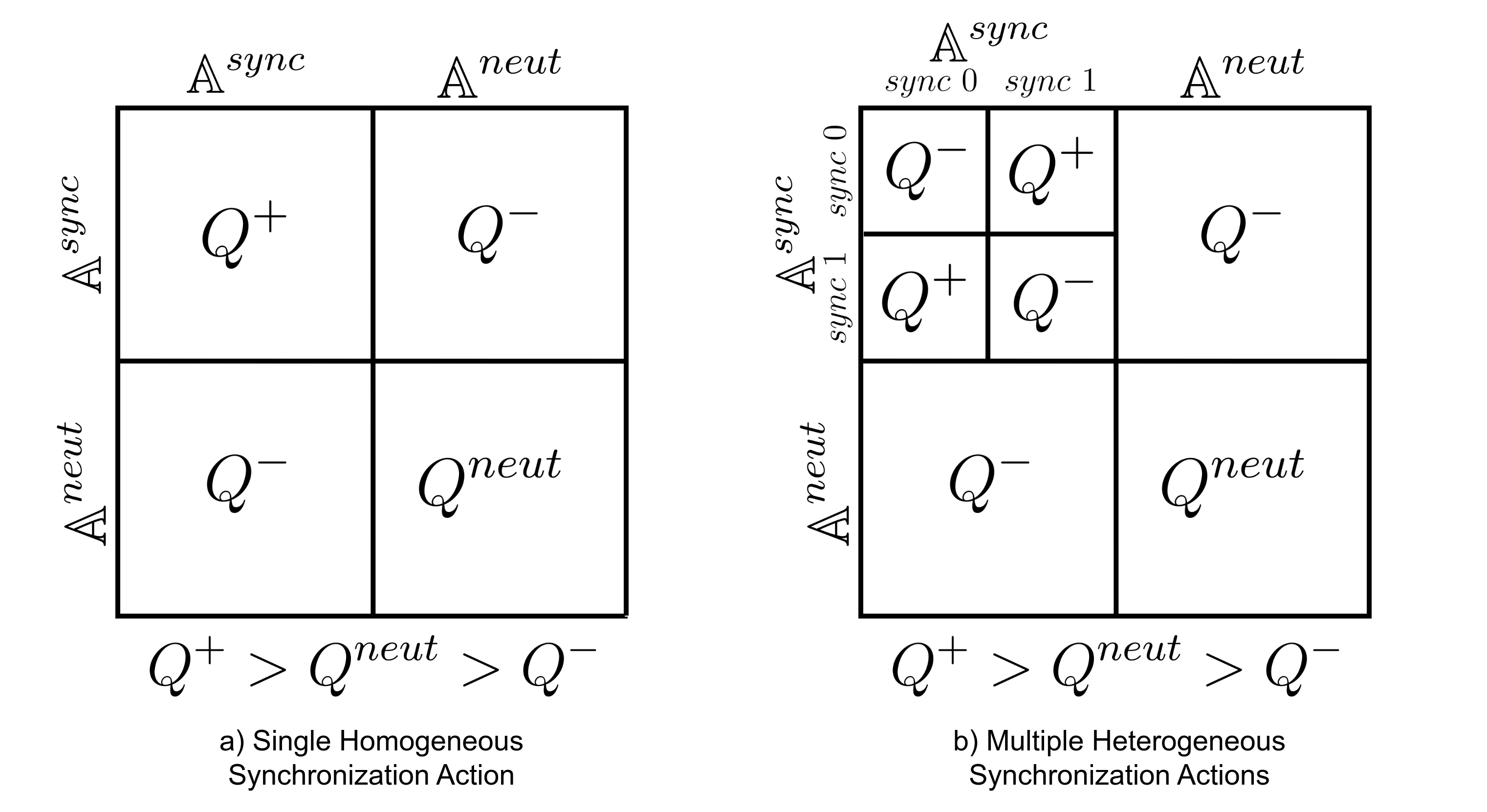}
  \caption{Visual representation of the payoff relationship in the Synchronized Predator-Prey Task for 2-agent sub-teams. Of note are the strict inequalities between the possible $Q$-values.}
  \label{fig:payoff}
\end{figure}

\textbf{States.}
The \textit{Synchronized Predator-Prey} environment is a discrete 10 x 10 grid with 8 predators and 8 prey. However, when considering sub-teams of three predators, the number of predators increases to 9. An episode is initialized with all agents randomly placed on the game grid. An episode ends once all predator sub-teams (i.e., number of predators required to capture a prey) have captured a prey or after 200 timesteps have passed.

\textbf{Transition Function.}
Agents are allowed to move one grid space per timestep. Moreover, predators can either select a capture action or a movement action (not both in the same timestep). Once a sub-team of predators have successfully captured a prey, they are all removed from the game grid for the remainder of the episode. Each agent occupies one grid square, and no agents may pass through each other. The predator agents' actions are controlled by an RL policy. While the prey agents' randomly select their movement from available open positions, or they remain still if no open surrounding position is available. Unattainable actions, such as moving into an occupied grid position, capturing without being adjacent to a prey, or crossing the grid boundary, are considered unavailable.

\textbf{Actions.}
All agents have the following five actions: \{remain still, left, right, down, up\}, which are considered the \textit{neutral actions} (${A}^{i, \text{ } neut}$). Additionally, predators have a sixth action, a homogeneous capture action for capturing prey agents. To capture, predators adjacent to a prey must synchronize their capture actions within the same timestep, which is the \textit{synchronization action} (${A}^{i, \text{ } sync}$). 

\textbf{Observations.}
The observation of a predator consists of a 5 × 5 sub-grid centered around itself, with one channel showing other predators represented by their id and another indicating prey. The observation also includes the position of the predator (itself) on the grid. If a predator has been removed after a successful capture they receive an observation of all zeros for the remainder of an episode.

\textbf{Reward.}
A successful capture of a prey is rewarded with a +10, but unsuccessful attempts by less than the required sub-team size are penalized with a -2 miscapture penalty. The maximum reward is dependent upon the total number of possible sub-teams (e.g., 40 for four 2-agent sub-teams and 30 for three 3-agent sub-teams). The reward is shared by all the predators.

\textbf{Heterogeneous Synchronized Predator-Prey.} A variant of Synchronized Predator-Prey introduces heterogeneous capture actions. For the heterogeneous variant, predators adjacent to the prey must synchronize \textbf{unique} capture actions within the same timestep. The number of available capture actions is equal to the number of predators required to capture a prey (i.e., 2 for sub-teams of two and 3 for sub-teams of three), which are the \textit{synchronization actions} (${A}^{i, \text{ } sync}$). 

We now show that Synchronized Predator-Prey satisfies the MST definition presented in Section \ref{sec:msts_def}. Consider the system states shown in Figure \ref{fig:synchronized_pred-prey_w_3_catcher}. In each case, $\mathbb{A}^+(s)$ is non-empty because there are two sub-teams of agents for which the execution of synchronization actions (as defined above) is optimal, as this would allow for the capture of the prey. Additionally, $\mathbb{A}^-(s)$ is non-empty because there are possible scenarios in which one agent might execute a synchronization action while the others do not, and doing so would lead to a large penalty, which is worse than if all the agents had executed neutral actions (as defined above). Therefore, the Synchronized Predator-Prey task meets the requirements of an MST as defined in Section \ref{sec:msts_def}. 

Figure \ref{fig:payoff} provides a visual representation of the payoff relationship when considering 2-agent sub-teams when they are in a state where successful synchronization is possible. Figure \ref{fig:payoff}a shows the achievable $Q$-values dependent on the actions the agents select when they are using a single homogeneous capture action. Figure \ref{fig:payoff}b shows the achievable $Q$-values dependent on the actions the agents select when they are using heterogeneous capture actions.

\begin{figure}[t]
  \centering
  \includegraphics[width=0.95\linewidth]{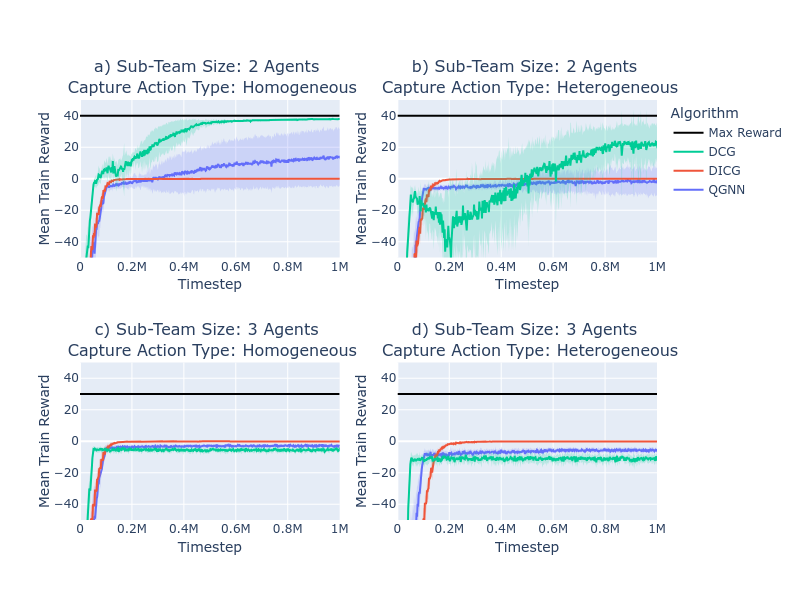}
  \caption{Train episode reward (Mean and shaded Standard Deviation) for ten independently trained models. The \textit{Full} CG topology was used, except for DICG which used an \textit{Attention} mechanism to create the graph. Note that SOTA methods did not solve MSTs well. Currently, DCG is the best solution (i.e., (a) and (b)) but cannot scale to larger sub-teams or handle more complex coordination (i.e., (c) and (d)).}
  \label{fig:penalty-full-all_conditions}
\end{figure}

\begin{figure}[h]
  \centering
  \includegraphics[width=0.95\linewidth]{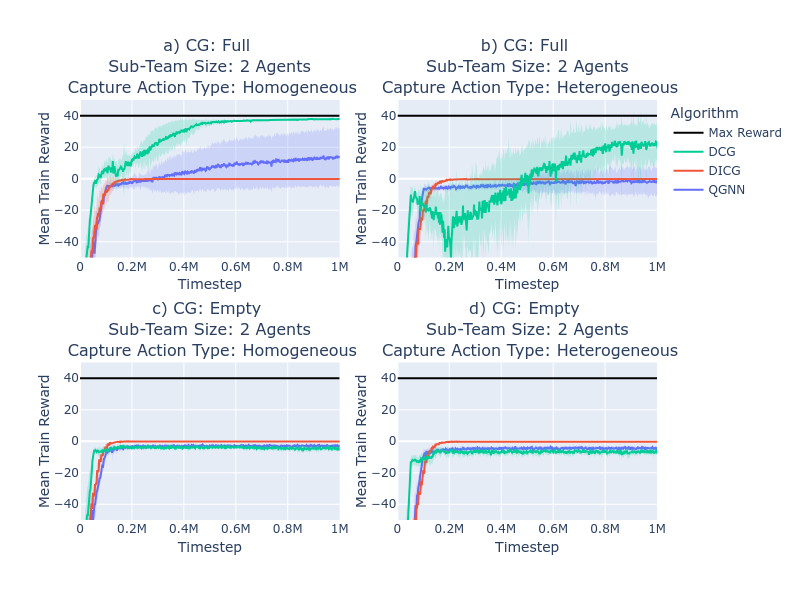}
  \caption{Train episode reward (Mean and shaded Standard Deviation) for ten independently trained models. The \textit{Full} CG topology was used for (a) and (b), except for DICG which used an \textit{Attention} mechanism to create the graph. For (c) and (d), the \textit{Empty} CG topology was used for all algorithms. Note that communication is necessary for MSTs but is not sufficient when complexity increases.}
  \label{fig:penalty-full_v_empty-2_agents}
\end{figure}

\section{Experiments}

To evaluate the algorithms introduced in Section \ref{related_work} in MSTs, we conducted experiments using the \textit{Synchronized Predator-Prey} task in four different configurations. The experimental configurations are as follows: (1) 2-agent sub-teams with homogeneous capture action, (2) 2-agent sub-teams with heterogeneous capture actions, (3) 3-predator sub-teams with homogeneous capture action, and (4) 3-predator sub-teams with heterogeneous capture actions. In addition, we used the following Coordination Graph (CG) topologies for our evaluations: \textit{Full} - each agent has an edge connection to every other agent, \textit{Empty} - no edge connections, and \textit{Attention} - a special case for DICG where an attention network generates the coordination graph at each timestep given the state of the environment. 

Figure \ref{fig:penalty-full-all_conditions} presents the results of our experiments with \textit{Full} CG topology for DCG and QGNN, and \textit{Attention} used for DICG. In general, the results showed that current SOTA methods utilizing coordination graphs had limited success with MSTs (see Figure \ref{fig:penalty-full-all_conditions}a  \& \ref{fig:penalty-full-all_conditions}b). Out of all the approaches evaluated, DCG performed the best with MSTs. DCG showed strong performance with 2-agent sub-teams and homogeneous capture actions but had marginal performance with heterogeneous capture actions (see Figure \ref{fig:penalty-full-all_conditions}a  \& \ref{fig:penalty-full-all_conditions}b). While QGNN showed marginal performance in the 2-agent sub-team with homogeneous capture actions condition (see Figure \ref{fig:penalty-full-all_conditions}a), it failed in the the other conditions Figure \ref{fig:penalty-full-all_conditions}b-\ref{fig:penalty-full-all_conditions}d). Note these results differ from those presented in DICG \cite{li2021dicg} and QGNN \cite{kortvelesy2022qgnn} as they were not using an MST for their experiments.

Figure \ref{fig:penalty-full_v_empty-2_agents} represents performance of the SOTA methods on MSTs with and without communication. For the experiments with communication, the \textit{Full} CG topology for DCG and QGNN, and \textit{Attention} used for DICG. In contrast, experiments without communication utilized the \textit{Empty} CG topology. The results provide empirical evidence that some form of communication is needed in order to solve an MST (compare Figure \ref{fig:penalty-full_v_empty-2_agents}a \& \ref{fig:penalty-full_v_empty-2_agents}b to \ref{fig:penalty-full_v_empty-2_agents}c \& \ref{fig:penalty-full_v_empty-2_agents}d). These results suggest communication is necessary to solve an MST. The communication method used must allow for agents to pass along the information that will allow them to correctly synchronize their actions.

\begin{figure}[t]
  \centering
  \includegraphics[width=0.95\linewidth]{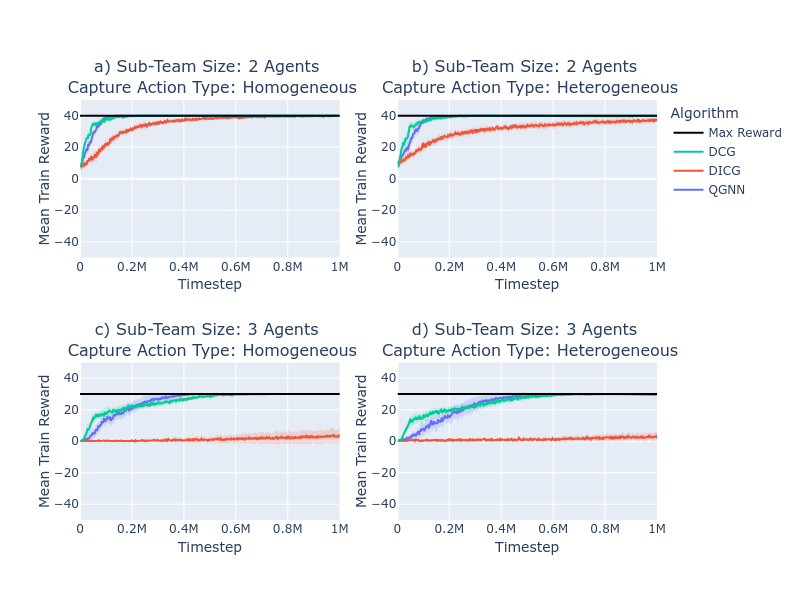}
  \caption{Train episode reward (Mean and shaded Standard Deviation) for ten independently trained models. The \textit{miscapture penalty} was disabled for these training iterations. The \textit{Full} CG topology was used, except for DICG which used an \textit{Attention} mechanism to create the graph. Note with the miscapture penalty disabled the task no longer satisfies the requirements for an MST and was solved by all SOTA methods.}
  \label{fig:no_penalty-full-all_conditions}
\end{figure}

Disabling the miscapture penalty was explored to confirm that the SOTA methods could solve a related non-MST task. It should be noted that the modified task is no longer an MST as defined in Section \ref{sec:msts_def}, as there is no difference in the $Q$-values when taking \textit{neutral actions} versus actions that would have been considered \textit{synchronization-negative} in the unmodified (MST) task. Figure \ref{fig:no_penalty-full-all_conditions} shows the results of our experiments using \textit{Full} CG topology for DCG and QGNN, and \textit{Attention} used for DICG. The data suggests that all considered SOTA methods solved the \textit{Synchronized Predator-Prey} task when not constrained by the requirements of an MST (see Figure \ref{fig:no_penalty-full-all_conditions}a-\ref{fig:no_penalty-full-all_conditions}d). However, DICG was unable to perform when the sub-team size was greater than two (see Figure \ref{fig:no_penalty-full-all_conditions}c \& \ref{fig:no_penalty-full-all_conditions}d). 

Together the results show the following: (1) DCG performs the best on MSTs, though it is unable to scale to more than 2-agent sub-teams or handle increased coordination complexity (i.e. heterogeneous capture actions), (2) communication is required to solve an MST, and (3) a penalty is critical to the definition of an MST.

\section{Discussion and Conclusion}

In this paper, we introduced and defined Multi-Agent Synchronization Tasks (MSTs), a novel multi-agent task subset aimed at assessing the capabilities of existing methods in addressing coordination challenges within multi-agent systems. Specifically, we introduced a distinctive Multi-Agent Synchronization Task termed "Synchronized Predator-Prey," designed to evaluate the effectiveness of prevailing techniques in handling complex coordination scenarios.

Our evaluation of SOTA methods on the Synchronized Predator-Prey task reveals that, currently, no approaches demonstrate robust performance in solving MSTs. Notably, DCG exhibits promising results (see Figure \ref{fig:penalty-full-all_conditions}a \& \ref{fig:penalty-full-all_conditions}b, albeit failing when coordination complexity increases (i.e. for 3-agent sub-teams and/or heterogeneous synchronization tasks - see Figure \ref{fig:penalty-full-all_conditions}c \& \ref{fig:penalty-full-all_conditions}d).

Three critical questions and our hypothesized solutions are presented here based on our experimental analysis. First, given the success of DCG in MSTs why does it fail to scale (i.e., to 3-agent sub-teams)? We believe DCG fails to scale mainly because it only models pairwise payoff factors. Thus, DCG cannot fully represent the coordination relationship between groups of more than 2-agents. Second, why does DCG struggle to handle heterogeneous capture actions? We currently hypothesize that DCG struggles with heterogeneous capture actions because the action synchronization facilitated by Max-Plus message passing \cite{kok2006MaxPlus} cannot fully address the need to coordinate unique actions between partners. Third, why do methods (DICG \& QGNN) utilizing GNNs for message passing fail to solve Multi-Agent Synchronization Tasks? Our hypothesis posits that a loss of representational capacity occurs with the transition from Max-Plus message passing to the GNN message passing. The lack of representative capacity may result from no longer having the payoff function or a failure of the generic GNN messaging function.

Finally, the results raise questions about the adaptability of GNNs in capturing the intricacies of complex coordination tasks and prompt further exploration into the underlying causes of their limitations in this context. Our ongoing research agenda includes delving into message passing frameworks to find a solution to the representational capacity problem we have identified with our newly defined multi-agent sub-domains we call MSTs. Overall, by asking these questions we aim to deepen our understanding of the challenges posed by MSTs and guide future developments in multi-agent coordination methodologies.






\newpage

\bibliographystyle{ACM-Reference-Format} 
\bibliography{references}


\end{document}